\documentclass[twocolumn,pra,a4paper,showpacs,floatfix,preprintnumbers,amsmath, amssymb,superscriptaddress]{revtex4}
\usepackage{amsmath}
\usepackage{graphicx}
\usepackage{subfigure}
\usepackage{psfrag}
\usepackage[centerlast]{caption}
\usepackage[dvips]{color}

\usepackage{graphicx}% Include figure files
\usepackage{dcolumn}% Align table columns on decimal point
\usepackage{bm}% bold math
\usepackage{units}%units


\begin{document}

\title{Current-Phase Relation of a Bose-Einstein Condensate \\Flowing Through a Weak Link}

\author{F. Piazza}\affiliation{CNR-INFM BEC center and Dipartimento di Fisica, Universit\`a di Trento, I-38050 Povo, Italy}
\author{L. A. Collins}\affiliation{Theoretical Division, Mail Stop
B214, Los Alamos National Laboratory, Los Alamos, New Mexico 87545}
\author{A. Smerzi}\affiliation{CNR-INFM BEC center and Dipartimento di Fisica, Universit\`a di Trento, I-38050 Povo, Italy}

\date{\today}
                      
\begin{abstract}
We study the current-phase relation of a Bose-Einstein condensate flowing through a repulsive square barrier by solving analytically the one dimensional Gross-Pitaevskii equation. The barrier height and width fix the current-phase relation $j(\delta\phi)$, which tends to $j\sim\cos(\delta\phi/2)$ for weak barriers and to the Josephson sinusoidal relation $j\sim\sin(\delta\phi)$ for strong barriers. Between these two limits, the current-phase relation depends on the barrier width. 
In particular, for wide enough barriers, we observe two families of multivalued current-phase relations. Diagrams belonging to the first family, already known in the literature, can have % a positive slope of the current when the phase difference is close to $\pi$
two different positive values of the current at the same phase difference. The second family, new to our knowledge, % shows a negative slope of the current close to $\pi$.
can instead allow for three different positive currents still corresponding to the same phase difference.
Finally, we show that the multivalued behavior arises from the competition between hydrodynamic and nonlinear-dispersive components of the flow, the latter due to the presence of a soliton inside the barrier region.
%differing from the known reentrant diagram with phase differences across the system larger than $\pi$. 
\end{abstract}

\maketitle

\section{Introduction}
The current-phase relation characterizes the flow of a superfluid/superconductor through a weak link \cite{barone, likharev, packard_RMP}. The latter is a constricted flow configuration that can be realized in different ways:  i) apertures in impenetrable walls mostly for helium,  ii) sandwich or bridge structures for superconductors, and iii) penetrable barriers created by laser beams for ultracold dilute gases. 
% In general, a weak link is a device where pure tunneling conduction, giving rise to an ideal Josephson coupling between two bulks of material, can be accompanied by superfluid and/or normal conduction. 
Much information about such systems can be extracted from the current-phase relation, which, given a fluid, depends only on the link properties. For instance, with $^4${\rm He}, the transition from the usual AC Josephson effect to a quantized phase slippage regime \cite{varoquaux} corresponds to the switching from a sine-like current phase relation to a multivalued one \cite{packard}.

A weak link configuration can be modelled very generally upon taking a portion of a superfluid/superconductor
to have ``different conduction properties'' with respect to the rest of the system.  Two pieces of superconductor joined by a third superconducting region with a smaller coherence length provide one example, whose current-phase relation in one dimension has been studied with the Ginzburg-Landau equation \cite{baratoff}.

In the context of ultracold dilute gases, raising a repulsive penetrable barrier across the flow yields an equivalent configuration. For instance, with Bose-Einstein condensates (BEC), Josephson effect(s) have been theoretically studied \cite{smerzi,milburn,leggett,bergeman,chiofalo} and experimentally demonstrated using multiple well traps \cite{inguscio,oberthaler,steinhauer}.
Theoretically, the current-phase relation has been studied for a flow through a repulsive square well with fermions across the BCS-BEC crossover by means of one dimensional Bogoliubov-de Gennes equations \cite{strinati}, for weak barriers with bosons in a local density approximation \cite{watanabe}, and for fermions on the BEC side of the crossover using a nonlinear Schr\"odinger equation approach \cite{ancilotto}.

In this manuscript, we study the current-phase relation for a BEC flowing through a repulsive square well. The weak link configuration, and in turn the current-phase relation, is then determined by the barrier height with respect to the chemical potential and by the barrier width with respect to the healing length. Though we solve a one-dimensional Gross-Pitaevskii equation%  with this simplified choice of the weak link
, % we basically capture the essential features of the current-phase relation. Therefore,
the results presented in this manuscript are not just relevant for BECs, but also include the essential features of current-phase relations of superconducting or superfluid {\rm He}-based weak links when governed by the Ginzburg-Landau equation.
For any barrier width, we find that in the limit of zero barrier height, the current phase relation tends to  $j(\delta\phi)=c_{\infty}\cos(\delta\phi/2)$, with $c_{\infty}$ being the bulk sound velocity, which corresponds to the phase across a grey soliton at rest with respect to the barrier. On the other hand, if the barrier height is above the bulk chemical potential at zero current, the limit of tunneling flow is reached either when the barrier height is much bigger than the bulk chemical potential at zero current or when the barrier width is much larger than the bulk healing length. In this regime, we recover the the usual Josephson sinusoidal current-phase relation and obtain an analytical expression for the Josephson critical current as a function of the weak link parameters.
For barriers wider than the healing lenght inside the barrier region, we observe two families of multivalued (often called reentrant) current-phase relations. The first, already studied since the early works on superconductivity \cite{likharev, baratoff}, shows a positive slope of the current when the phase difference is close to $\pi$, thereby reaching a phase difference larger than $\pi$ at least for small currents. The second family, appearing at a smaller barrier height, has instead a negative slope of the current close to $\pi$, and in some cases can remain within the $0-\pi$ interval across the whole range of currents. These two families can also be distinguished by the maximum number of different positive currents corresponding to the same phase difference: two for the first family, three for the second one. As the first kind of reentrant behavior was proven to be connected to the onset of phase-slippage in the AC Josephson effect \cite{packard}, the second might then be connected to the appearance of new features in the Josephson dynamics.
We finally observe that the hysteresis characterizing both families of reentrant current-phase relations is always due to the competition between a hydrodynamic component of the flow and a nonlinear-dispersive component, the latter due to the presence of a soliton inside the barrier region. The two components can coexist only for barriers wide enough to accomodate a soliton inside. In this spirit, we develop a simple analytical model which describes very well reentrant regimes of current-phase relations.

\section{Theoretical Developments}

\subsection{ Stationary Solutions}
We consider a dilute repulsive Bose-Einstein condensate at zero temperature flowing through a 1D rectangular potential barrier. We look for stationary solutions of the 1D GPE \cite{dalfovo}:
\begin{equation}
- \frac{\hbar^2}{2m}\partial_{xx}\Psi + V_{\rm ext}(x) \Psi +
g | \Psi |^2 \Psi = \mu \Psi ,
\label{eq:gp}
\end{equation}
where $\Psi(x)=\sqrt{n(x)}\exp[i\phi(x)]$ is the complex order
parameter of the condensate, $\mu$ is
the chemical potential, and $g=4\pi\hbar^2a_s/m$ with $m$ the atom mass and $a_s>0$
 the $s$-wave scattering length. 
The order parameter phase $\phi(x)$ is related to the superfluid velocity via $v(x)=(\hbar/m)\partial_x \phi(x)$.
The piecewise constant external potential describes the rectangular barrier of width $2d$ and height $V_0$:
\begin{equation}
\left\{
\begin{aligned}
V_{\rm ext}(x)&=V_0&,\ |x|\leq d\ ,\\
V_{\rm ext}(x)&=0&,\ |x|>d\ .
\end{aligned}
\right.
\label{eq:barrier}
\end{equation} 
We consider solutions of Eq.~(\ref{eq:gp}) which are symmetric with respect to the point $x=0$, therefore discarding cases in which a reflected wave is present \cite{leboeuf}. Such symmetric solutions in the presence of a barrier exist due to the nonlinearity in the GPE.
We also restrict our analysis to subsonic flows $v_{\infty}\leq c_{\infty}$, with $c_{\infty}=\sqrt{gn_{\infty}/m}$ being the sound velocity for a uniform condensate of density $n_{\infty}$.
As boundary conditions, we fix the condensate density $n_{\infty}$ and velocity $v_{\infty}$ at $x=\pm\infty$, thereby determining the chemical potential $\mu=g n_{\infty}+\frac{1}{2}m v_{\infty}^2$. 
Using the relation $\Psi=\sqrt{n}\exp[i\phi]$,  Eq.~(\ref{eq:gp}) can be split into a continuity equation, stating the constancy in space of the current $j=n(x)v(x)=n_{\infty}v_{\infty}$, and an equation for the density only
\begin{equation}
\mu=-\frac{\hbar^2}{2m}\frac{\partial_{xx}\sqrt{n}}{\sqrt{n}}+\frac{mj^2}{2n}+V_{\rm ext}(x)+gn\ ,
\label{eq:denequation}
\end{equation}
where we have used the continuity equation $v(x)=j/n(x)$.
Its solution $n(x)$ is expressed in terms of Jacobi elliptic functions \cite{mamaladze, baratoff}. Due to symmetry, we need only consider half of the space $x>0$. The solution outside the barrier $x>d$ becomes
\begin{equation}
n_{{\rm out}}(x)=n_{\infty}-A_{\infty}+A_{\infty}{\rm tanh}^2\left[\sqrt{\frac{mg}{\hbar^2}A_{\infty}}(x-d)+x_0\right]\ ,
\label{eq:solutionout}
\end{equation}
where $x_0={\rm arctanh}\sqrt{(n_{d}-mv_{\infty}^2/g)/A_{\infty}}$, $A_{\infty}=n_{\infty}-mv_{\infty}^2/g\geq 0$, and $n_d$ is the density at the barrier edge $x=d$.
The solution inside the barrier $x<d$ is:
\begin{equation}
\left\{
\begin{aligned}
n_1(x)&=n_0+A_1\frac{{\rm sn}^2[b_1x,k_1]}{{\rm cn}^2[b_1 x,k_1]}&,&\ \Delta\geq 0\text{ and } A_1\geq 0\ ,\\
n_2(x)&=n_0+A_2\frac{1-{\rm cn}[b_2x,k_2]}{1+{\rm cn}[b_2 x,k_2]}&,&\ \text{else}\ ,
\end{aligned}
\right.
\label{eq:solutions}
\end{equation} 
where $n_0$ is the density at $x=0$, $\Delta=(n_0-2\tilde{\mu}/g)^2-4mj^2/gn_0$ with $\tilde{\mu}=\mu-V_0$, $A_1=3n_0/2-\tilde{\mu}/g-\sqrt{\Delta}/2$, and $A_2=\sqrt{2(n_0^2-n_0\tilde{\mu}/g+mj^2/2gn_0)}$. Finally, the parameters entering the Jacobi sines ${\rm sn}$ and cosines ${\rm cn}$ are $b_1=\sqrt{mg(\sqrt{\Delta}+A_1)/\hbar^2}$, $k_1=(\sqrt{\Delta}/(\sqrt{\Delta}+A_1))^{1/2}$, and $b_2=\sqrt{4mgA_2/\hbar^2}$, $k_2=((A_2-A_1-\sqrt{\Delta}/2)/2A_2)^{1/2}$.
Given $n_{\infty}$ and $v_{\infty}$, we are left with two free parameters: $n_0$ and $n_d$, which are next determined by matching the density and its derivative at the barrier edge $x=d$. 
First, the derivative matching condition, enforced using the first integral of Eq.~(\ref{eq:denequation}), allows us to write $n_d$ as a function of $n_0$ for any value of $\Delta$:
\begin{equation}
n_d=\frac{g}{2V_0}\left[n_{\infty}^2+n_0^2+\frac{mv_{\infty}^2n_{\infty}}{g}(2-\frac{n_{\infty}}{n_0})-2\frac{\tilde{\mu}}{g}n_0\right]\ .
\label{eq:nd}
\end{equation}
The density matching equation $n_d=n_{i}(d)$ ($i=1,2$) is then solved by a numerical root finding method, yelding $n_0$. 

Two bounded solutions are always found \cite{solutionkind}, an example is given in the upper panel of Fig.~\ref{fig1}. In the following, the solution which tends to a plane wave for $V_0\to 0$ will be referred to as ``upper solution'', while the one tending to a grey soliton will be called ``lower solution'' \cite{stability}. For given barrier parameters $V_0$ and $d$, and at a fixed density at infinity $n_{\infty}$, the solutions exist up to a critical injected velocity $v_{\infty}=v_c<c_{\infty}$, at which they merge and disappear. This behavior was found in the case of a repulsive delta potential in \cite{hakim}. Similarly, in \cite{leboeuf}, the same kind of merging was reported for a 1D BEC flow through a repulsive square well when the width of the latter increases.
% The study of the stability of the two solutions is beyond the scope of the present paper. This issue has been addressed in \cite{hakim}. One solution was shown to be unstable while the other was regarded as stable. 

\subsection{ Current-phase Relation}
\begin{figure}[]
\hspace{0mm}
\includegraphics[scale=.26]{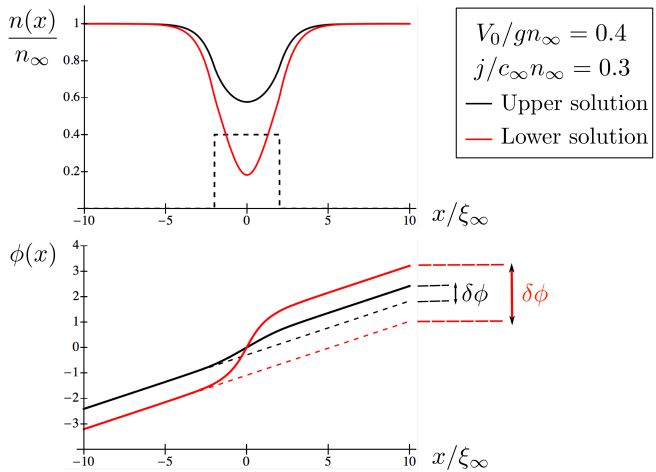}
\caption{ (Color online) Typical solutions for a barrier with width $2d=4\xi_{\infty}$% , $V_0=0.4gn_{\infty}$, and $j=0.3c_{\infty}n_{\infty}$
  . Density (upper panel) and phase (lower panel) as a function of position are shown for both the upper and the lower solution (see inset). Dashed lines in the lower panel correspond to the phase accumulated by a plane wave in absence of barrier.} \label{fig1}
\vspace{-0.0in}
\end{figure}

As pointed out in the Introduction, the current-phase relation $j(\delta\phi)$ for a given superfluid only depends on the properties of the weak link, in our case the barrier height $V_0$ with respect to the chemical potential and the width $2d$ with respect to the healing length. For a fixed current $j=v(x)n(x)$, the phase difference across the system is calculated using the relation $\phi(x)=\int^{x}dy(m/\hbar)j/n(y)$ and then renormalized by the phase accumulated by the a plane wave with the same boundary conditions in absence of barrier (see lower panel in Fig.~\ref{fig1}). Two different values of $\delta\phi$ are found, corresponding to the upper and lower solutions.

In this section, we will use dimensionless quantities, employing the chemical potential at zero current $gn_{\infty}$ as the unit of energy, the bulk healing length $\xi_{\infty}=\hbar/\sqrt{2mgn_{\infty}}$ as the unit of length, and $\hbar/gn_{\infty}$ as the unit of time.  
Exploiting the symmetry of the system about $x=0$, the phase difference can be written as $\delta\phi_{i}=\lim_{\substack{x\to\infty}}[\int_{0}^{d}\!\!jdy/n_{i}(y)+\int_{d}^{x}\!\!jdy/n_{out}(y)-jx/n_{\infty}]\ ,i=1,2$, where the third term is the phase difference accumulated by the plane wave. The limit can be calculated using Eq.~(\ref{eq:solutionout}), yielding
\begin{equation}
\delta\phi_{i}=\!\!\int_{0}^{d}\!\!\frac{jdy}{n_{i}(y)}-jd+2\!\left[\arcsin(\frac{j}{\sqrt{2n_i(d)}})-\arcsin(\frac{j}{\sqrt{2}})\right].
\label{eq:phase2}
\end{equation}
The first two terms in Eq.~(\ref{eq:phase2}) correspond to the phase acquired inside the barrier while the third, which we can call the pre-bulk term, gives the phase accumulated outside the barrier, where the density has not yet reached its bulk value $n_{\infty}$. We have taken the latter to be one.

\begin{figure}[]
\hspace{0mm}
\includegraphics[scale=.26]{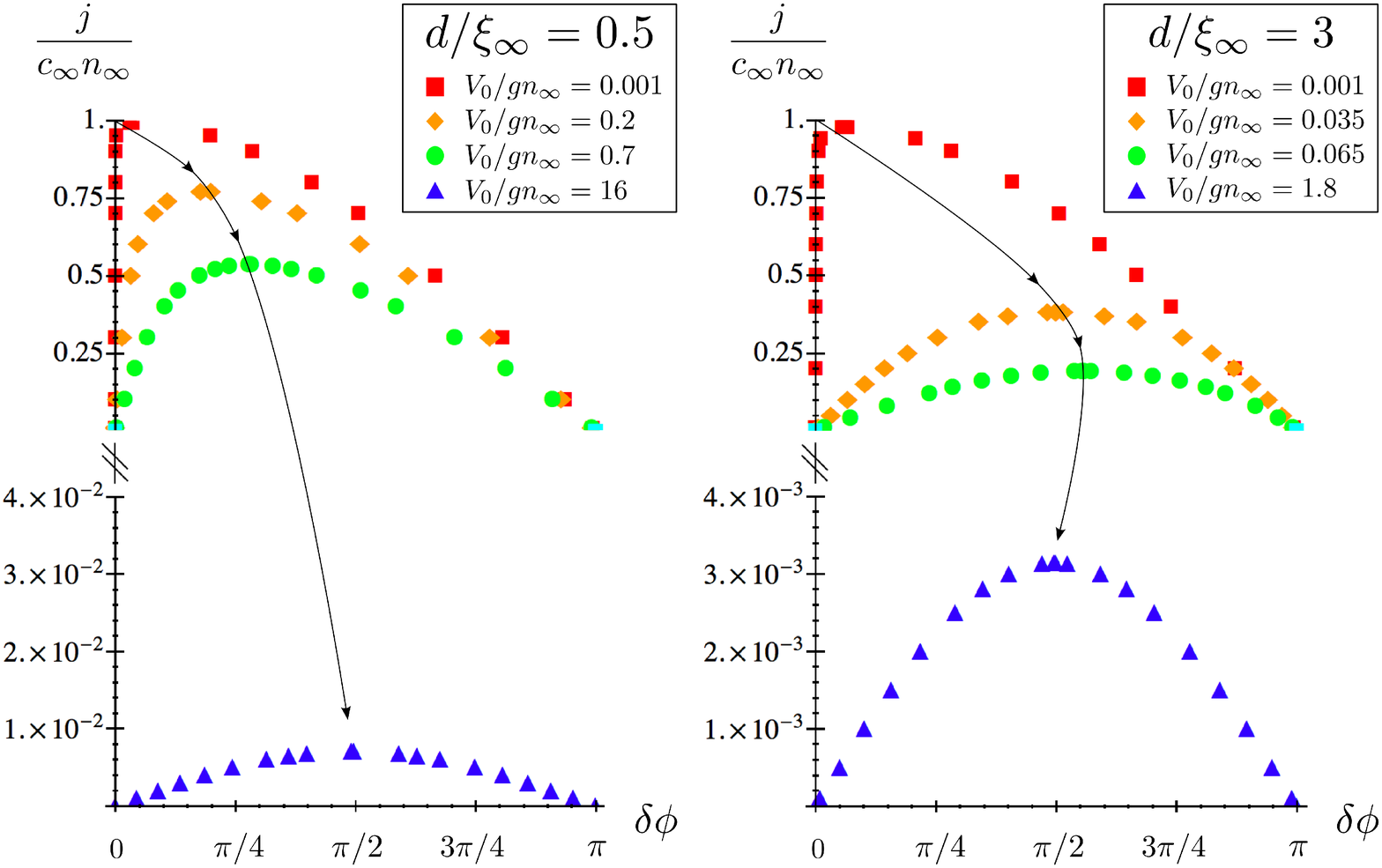}
\caption{ (Color online) Current-phase relation for a barrier with half width $d=0.5\xi_{\infty}$ (left panel) and $d=3\xi_{\infty}$ (right panel), for different barrier heights $V_0$ (see inset). The arrows sketch the behavior of the maximum of the curves upon increasing $V_0$.} \label{fig2}
\vspace{-0.0in}
\end{figure}

In Fig.~\ref{fig2}, the current phase relation is shown for different barrier widths and heights. Each curve has a maximum at the point $(\delta\phi_c,j_c)$, with $j_c=n_{\infty}v_c$ being the critical current at which the two stationary solutions merge and disappear. The upper solutions constitute the part of the current-phase diagram which connects the maximum with the point $(\delta\phi=0, j=0)$, while the lower ones belong to the branch connecting the maximum to the point $(\delta\phi=\pi, j=0)$. 
Indeed, we will now show that, for any $d$ and $V_0\to 0$, the upper branch tends to a plane wave, while the lower branch tends to a grey soliton. In order to have a finite $n_d$ in this limit, the term in square brackets in Eq.~(\ref{eq:nd}) must tend to zero, yielding a cubic equation with two coincident solutions $n_0=1$ and a third $n_0=j^2/2$, where we have set $n_{\infty}=1$ for simplicity. For $n_0=1$ we have $\Delta=(1-j^2/2)^2\geq 0$, corresponding to the plane wave solution $n_1(x)=n_0=1$. For $n_0=j^2/2$ we obtain instead $\Delta=0$, $A_1=j^2/2-1<0$, corresponding to a grey soliton solution $n_2(x)=n_0+(1-n_0)\tanh^2(\sqrt{1/2-n_0/2}x)$.
Therefore, in this limit $\delta\phi_1=0$ for any $j$, meaning that the upper branch is actually a vertical line, while for the lower branch we have
\begin{equation}
\cos(\frac{\delta\phi}{2})=\frac{j}{\sqrt{2}}.
\label{eq:freephase}
\end{equation} 
This curve has a maximum at $\delta\phi_2=0$, corresponding to  $j=\sqrt{2}$, that is, the sound velocity $c_{\infty}$ in dimensionless units. 

The arrows in Fig.~\ref{fig2} sketch the behavior of the maximum of the current-phase relation $(\delta\phi_c,j_c)$ as the height $V_0$ is increased at a fixed barrier width $2d$. For any width, the current-phase diagram initially takes a cosine-like shape (Eq.~(\ref{eq:freephase})) when $V_0\sim 0$ (red squares in Fig.~\ref{fig2}), and tends to a $\sin(\delta\phi)$ shape for sufficiently large $V_0$, characterizing the Josephson regime of tunneling flow  (blue triangles in Fig.~\ref{fig2}). Between these two limits, the behavior of the maximum is determined by the barrier width. For thin barriers ($d\lesssim\xi_{\infty}$), as shown in the left panel of Fig.~\ref{fig2}, the point $(\delta\phi_c,j_c)$ moves down-right, reaching the Josephson regime keeping $\delta\phi_c$ always smaller than $\pi/2$. On the other hand, for sufficiently wider barriers, $\delta\phi_c$ is able to reach values larger than $\pi/2$ within a finite range of $V_0$ during the down-right displacement of the maximum. The latter then moves down-left to finally enter the Josephson regime, as shown in the right panel of Fig.~\ref{fig2}. We note that in this way, while $V_0$ is increased, the phase $\delta\phi_c$ takes the value $\pi/2$ twice, but only the second time entering  the Josephson regime with a sinusoidal current-phase relation. The first time (orange diamonds in the right panel of Fig.~\ref{fig2}) the flow is not yet in the tunneling regime since $V_0$ is much smaller than the chemical potential, indeed the current-phase relation is symmetrical with respect to $\pi/2$, but not sinusoidal. 

\subsection{Josephson Regime}
As described in the previous section, for strong enough barriers the flow enters the tunneling regime, and the current-phase relation takes a sinusoidal form. In the following, we will describe  analytically this behavior, deriving a relation between the Josephson critical current and the barrier parameters $V_0$ and $d$.
Since we are now interested in tunneling flows, we will take $V_0>gn_{\infty}$ and will show that the Josephson regime is attained by either increasing the barrier height $V_0$ or its width $2d$.

Since in this regime the injected velocity of a stationary flow $v_{\infty}$ must be much smaller than the sound velocity $c_{\infty}$, the chemical potential can be written as $\tilde{\mu}\simeq gn_{\infty}-V_0<0$, upon neglecting the kinetic energy term $mv_{\infty}^2/2$. Moreover, the density inside the barrier $n_0$ being exponentially small, we can write $\Delta\simeq (2\tilde{\mu}/g)^2-s>0$, with $s=4(n_0\tilde{\mu}+mj^2/n_0)/g$, where we have neglected $n_0^2$. Thus, both the upper and lower solutions are of the kind $n_1(x)$, with $A_1\simeq n_0-mj^2/2n_0\tilde{\mu}$, $b_1\simeq\sqrt{2m|\tilde{\mu}|/\hbar^2}$, and $k\simeq 1$. The density in the Josephson regime has thus the form
\begin{equation}
n_{{\rm jos}}(x)=n_0+(n_0+\frac{mj^2}{2n_0|\tilde{\mu}|})\sinh^2\left(\sqrt{\frac{2m}{\hbar^2}|\tilde{\mu}|}\ x\right).
\label{eq:denjos}
\end{equation}
In order to write the density matching equation $n_d=n_1(d)$, we approximate $n_d$ by discarding both $n_0^2$ and $2mv_{\infty}^2n_{\infty}/g$ in Eq.~(\ref{eq:nd}), obtaining a quadratic equation for $n_0$. % : 
% \begin{equation}
% n_0^2\left(1+\frac{\tilde{\mu}}{V_0}+\sinh^2\sqrt{\frac{|\tilde{\mu}|}{\epsilon_{d}}}\right)-n_0\frac{gn_{\infty}^2}{2V_0}+mj^2\left(\frac{1}{2V_0}-\frac{\sinh^2\sqrt{\frac{|\tilde{\mu}|}{\epsilon_{d}}}}{2\tilde{\mu}}\right),
% \label{eq:matchjos}
% \end{equation}
Further assuming that $\sinh^2\sqrt{\frac{|\tilde{\mu}|}{\epsilon_{d}}}\gg 1$, with $\epsilon_d=\hbar^2/2md$ being the kinetic energy associated with the barrier length scale $d$, the solutions of the above equation are of the form $n_{0}^{+/-}=\bar{n}_0(1\pm\sqrt{1-q})$ with $\bar{n}_0=(gn_{\infty}/4V_0)n_{\infty}/\sinh^2\sqrt{|\tilde{\mu}|/\epsilon_{d}}$, and $q=mj^2/2\bar{n}_0^2|\tilde{\mu}|$.

The Josephson critical current corresponds to the merging of the two solutions at $q=1$, and reads
\begin{equation}
j_{\rm jos}=n_{\infty}\sqrt{\frac{2|\tilde{\mu}|}{m}}\frac{gn_{\infty}}{V_0}{\rm e}^{-2\sqrt{|\tilde{\mu}|/\epsilon_{d}}},
\label{joscrit}
\end{equation}
where we have used  $\sinh(x)\simeq\exp(x)/2$, for $x\gg 1$. The critical velocity for a bosonic and fermionic superfluid flowing through a repulsive square well has been calculated in \cite{watanabe} within the local density approximation (for BEC case see also \cite{lesz}). Analytical expressions for the critical current of a BEC flow were found for both slowly varying and weak barriers \cite{hakim}. Eq.~(\ref{joscrit}) thus enriches the above set of analytical results by providing the critical current for strong barriers.

Finally, we calculate the current-phase relation using Eq.~(\ref{eq:phase2}). Since $j_{\rm jos}$ is exponentially small, only the first term in Eq.~(\ref{eq:phase2}) contributes, and the integral can be performed analytically to obtain:
\begin{equation}
\delta\phi^+=\arcsin\left(\frac{j}{j_{\rm jos}}\right)\ ,\delta\phi^-=\pi-\arcsin\left(\frac{j}{j_{\rm jos}}\right).
\end{equation}
Thus, we recover the sinusoidal current-phase relation characterizing a Josephson regime of tunneling flow.

\begin{figure}[]
\hspace{0mm}
\includegraphics[scale=.25]{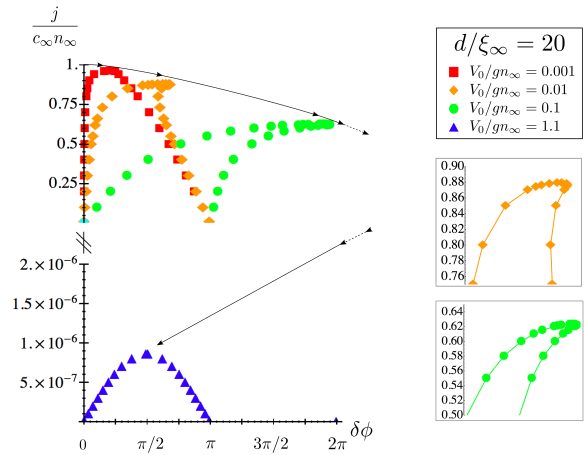}
\caption{ (Color online) Current-phase relation in the reentrant regime $d=20\xi_{\infty}$. Insets show in more detail the shape of the diagram close to the critical point. The arrows sketch the behavior of the maximum of the curves upon increasing $V_0$.} \label{fig3}
\vspace{-0.2in}
\end{figure}

\begin{figure}[]
\hspace{0mm}
\includegraphics[scale=.25]{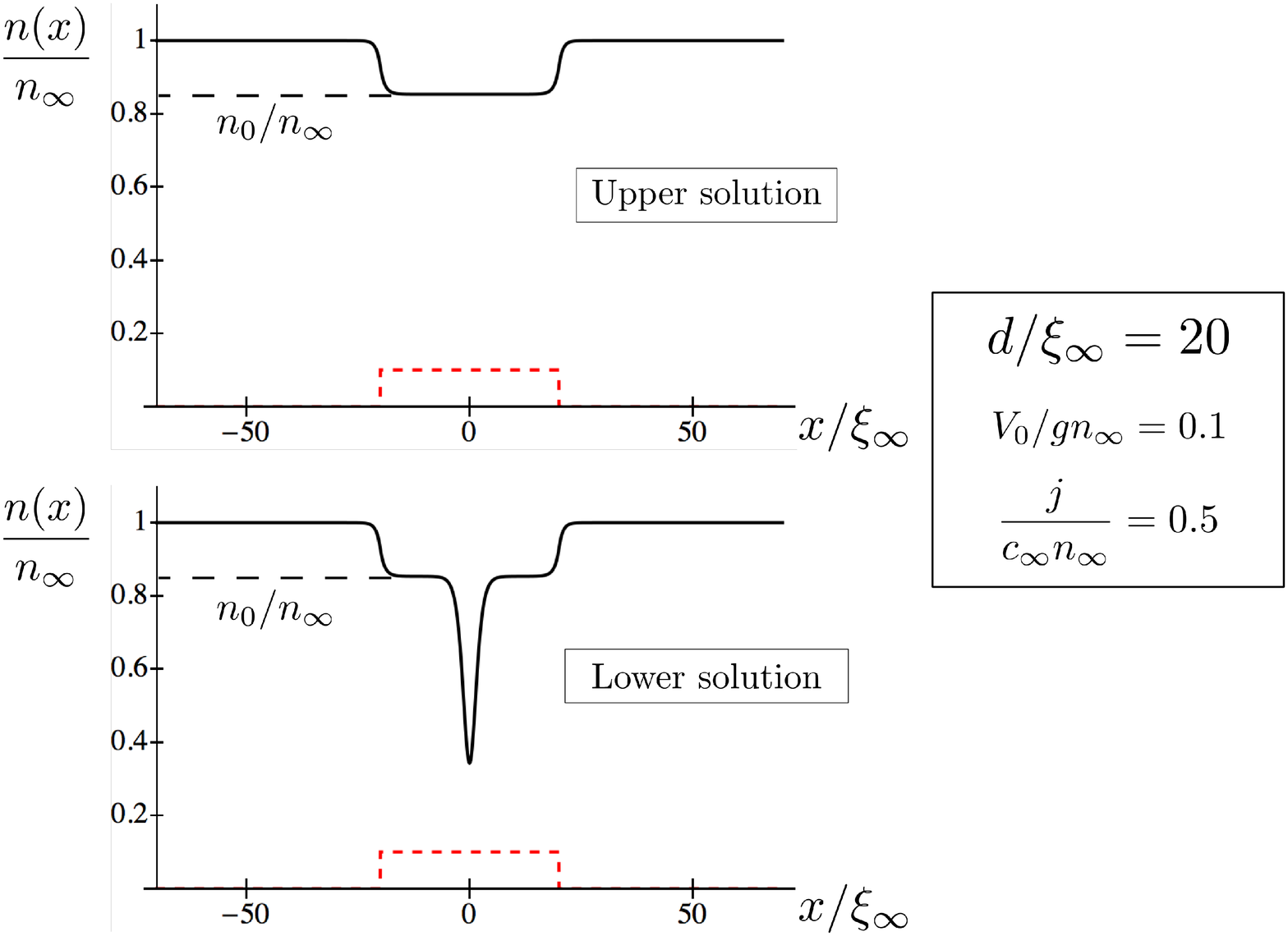}
\caption{ (Color online) Typical density profiles in the reentrant regime $d=20\xi_{\infty}$ for the upper and lower solutions.} \label{fig4}
\vspace{-0.2in}
\end{figure}
 
\subsection{ Reentrant Regimes}
When the barrier width $2d$ greatly exceeds the healing length, the current-phase relation becomes multivalued as shown in Fig.~\ref{fig3} for $d=20\xi_{\infty}$. These so-called reentrant current-phase diagrams were first predicted for long superconducting weak links \cite{likharev, baratoff}. Remarkably, they have been experimentally demonstrated with superfluid $^4{\rm He}$ \cite{packard}.

In our system, two kinds of multivalued diagrams are found for a fixed barrier width $2d$. We designate the first kind (green dots), appearing at larger barrier heights $V_0$, as reentrant type I, and the second kind (orange diamonds in Fig.~\ref{fig3}), occurring for smaller $V_0$,  as reentrant type II. They differ in the behavior of the lower branch at small currents $j\ll 1$. The phase decreases with increasing $j$ in type II diagrams while it increases in type I, reaching values larger than $\pi$ for $j\ll 1$.
These two families of diagrams also differ in the number of positive values of the current $j$ corresponding to the same phase difference $\delta\phi$. Indeed, diagrams of type I can have two values of $j$ at the same $\delta\phi$, while diagrams of type II can allow for three.
The existence of these type II diagrams, to our knowledge, has not been discussed in the literature. In \cite{packard}, only type I current-phase relations were observed and connected to the onset of phase-slip dissipation in the system. In this spirit, a type II diagram might for instance lead to an instability of a different kind. 

In the remainder of this section, we will develop an analytical model that captures the essential features underlying both kinds of reentrant current-phase diagrams. 
Examination of typical density profiles belonging to the reentrant regime (see Fig.~\ref{fig4}) suggests to construct the lower solution by starting from the upper one at the same current, then adding a grey soliton inside the barrier region.
Since we are dealing with wide barriers, we describe the upper solution in the local density approximation (LDA) (its current-phase relation in this approximation is also discussed in \cite{watanabe}). At a fixed current $j$, the density inside the barrier  $|x|<d$ is constant and equal to $n_0=\tilde{\mu}/3g+2\tilde{\mu}\cos(\omega/3)/3g$, when $j<j_{\rm th}$, or $n_0=\tilde{\mu}/3g+2\tilde{\mu}\cos(\pi/3-\omega/3)/3g$, when $j>j_{\rm th}$ with $j_{\rm th}$  defined by $j_{\rm th}^2=4\tilde{\mu}^3/27mg^2$, and $\omega=\arccos|1-27mg^2j^2/4\tilde{\mu}^3|$. The phase difference calculated within LDA misses the pre-bulk term in Eq.~(\ref{eq:phase2}), thus, for the upper branch, it is simply $\delta\phi_1=2mjd(1/n_0-1/n_{\infty})/\hbar$.
Now, for the density profile of the lower branch $n_2(x)$, we take a grey soliton (Eq.~(\ref{eq:solutionout})) placed inside the barrier at $x=0$, with a bulk density given by $n_0$ and a bulk velocity $v_0=j/n_0$ while in the region $|x|>d$ we keep the density profile of the upper branch, that is, a constant density $n_{\infty}$ and velocity $j/n_{\infty}$.
Notice that in this section $n_0$ stands for the density of the upper solution at $x=0$, as indicated in Fig.~\ref{fig4}. The density at $x=0$ for the lower solution corresponds to the center of the dip in the grey soliton density profile. 
Finally, using Eq.~(\ref{eq:phase2}) we obtain the phase difference for the lower branch
\begin{equation}
\delta\phi_2=\delta\phi_1+2\arccos\sqrt{\frac{mv_0^2}{gn_2(d)}}.
\label{reentrantphase}
\end{equation}
At a given current $j$, the overall phase difference corresponding to the lower solution has two contributions:  1) the ``hydrodynamics phase'' $\delta\phi_1$ coming from LDA and 2) the ``nonlinear-dispersive phase'' $\delta\phi_{\rm sol}=2\arccos\sqrt{mv_0^2/gn_2(d)}$ accumulated across the grey soliton. While $\delta\phi_1$ is a monotonically increasing function of $j$, $\delta\phi_{\rm sol}$ is instead monotonically decreasing, starting from $\pi$ at zero current \cite{deltaphisol}. 

Therefore, the hysteresis characterizing a reentrant current-phase relation is due to the competition between the hydrodynamic and the nonlinear-dispersive components of the flow, which can coexist only for barriers wide enough to accomodate a soliton inside. 
In particular, we can derive a condition for the appearance of type I reentrant behavior upon expanding Eq.~(\ref{reentrantphase}) for small currents, and requiring $\delta\phi_2>\pi$. Using $\arccos(x)\simeq\pi/2-x$, for $x\ll 1$, we get $\delta\phi_2\simeq\pi+2j\nu$, where $\nu=md(1/n_0-1/n_{\infty})/\hbar-\sqrt{m/gn_0^3}$, and we have taken $n_2(d)\simeq n_0$. The condition for type I reentrance to appear is thus $\nu>0$. For $j\ll 1$, we have $n_0\simeq\tilde{\mu}/g\simeq n_{\infty}-V_0/g$, and since within the LDA approximation $V_0\ll gn_{\infty}$, the condition $\nu>0$ takes the simple form
\begin{equation}
\frac{V_0}{gn_{\infty}}\ \frac{d}{\xi_0\sqrt{2}}>1,
\label{condreentrant}
\end{equation}
with $\xi_0=\hbar/\sqrt{2mgn_0}$ being the healing length inside the barrier region where the density is $n_0$.
Equation (\ref{condreentrant}), holding for $V_0\ll gn_{\infty}$, has a clear physical meaning: in order to have a type I reentrant current phase diagram, the barrier width $2d$ must be sufficiently larger than $2\sqrt{2}\xi_0$, which is the characteristic size of a soliton placed inside the barrier.
% The above equation also allows us to determine whether, for a fixed barrier width, the type II reentrant regime will appear at least for some height $V_0$. Since the l.h.s of Eq.~(\ref{condreentrant}) has a maximum at $V_0=2gn_{\infty}/3$ we have that, if:
% \begin{equation}
% \frac{d}{\xi_{\infty}}>(\frac{3}{2})^{3/2},
% \label{condreentrantbis}
% \end{equation}
% then type II reentrance will appear.

\begin{figure}[]
\hspace{0mm}
\includegraphics[scale=.25]{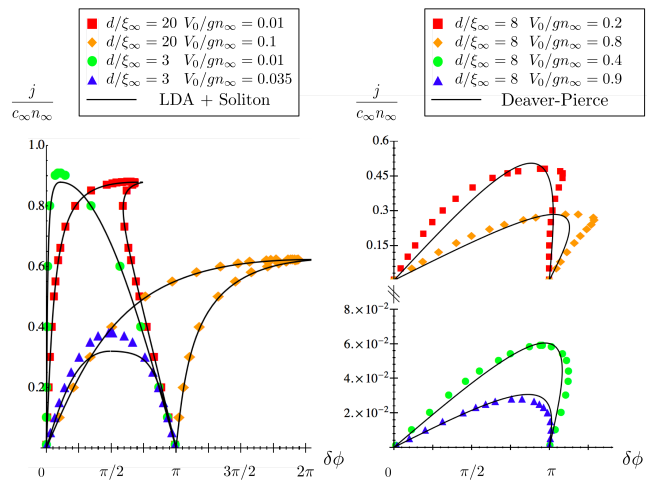}
\caption{ (Color online) Comparison between the LDA+soliton model (left panel) or the Deaver-Pierce model (right panel) and the exact results.} \label{fig5}
\vspace{-0.0in}
\end{figure}

In the left panel of Fig.~\ref{fig5}, we compare the current-phase relation calculated with the above model (solid lines) to the exact results. Within the reentrant regime, for both type I and type II, the agreement is striking with only  slight differences close to the the maximum $(\delta\phi_c,j_c)$. On the other hand, for thin/strong barriers, LDA, and in turn the above model, is in clear disagreement with the exact results. (See cases $d=3\xi_{\infty}$ in the left panel of Fig.~\ref{fig5}). 

The difference between type I and type II diagrams has a physical interpretation within the above model, namely that the hydrodynamic component of the flow dominates for all currents in type I reentrance (excluding the region $j\simeq j_c$ \cite{closetotip}), while it is overcome by the nonlinear-dispersive part for sufficiently small current in type II.

In the literature \cite{likharev}, multivalued current-phase relations are typically modelled by describing the weak-link with an equivalent circuit containing a linear inductance in series with a sinusoidal inductance, the latter corresponding to an ideal Josephson junction. 
When compared to our GPE exact results (see right panel in Fig.~\ref{fig5}), this model in general fails to describe the curvature of both branches, and misses type II reentrance, as well as nearly free regimes (e.g. red squares in Fig.~\ref{fig2},\ref{fig3}) since it does not allow the phase $\delta\phi_c$, corresponding to the maximum of the diagram, to be smaller than $\pi/2$. It proves quite accurate only for sufficiently large barrier heights $V_0$, very close to the Josephson regime. 
In the $^4${\rm He} experiment of \cite{packard}, this so called Deaver-Pierce model \cite{deaver}  agrees well with the measured current-phase relation. This might be due to the fact that these experiments are performed with a fixed weak link configuration, moving the system across the transition between Josephson and type I reentrant regime, upon changing the $^4${\rm He} healing length with temperature, but always staying sufficiently close to the Josephson regime.

\section{ Conclusions}
We studied the current-phase relation of a BEC flowing through a weak link created by a repulsive barrier. The link is thus modelled by two parameters, the barrier height and width, which fix the current-phase relation. Even though we solved a simplified model, we believe that the results obtained will also be relevant for superconducting and superfluid {\rm He}-based weak links.

We obtained analytical results for the weak barrier limit, for which the current-phase relation has a $\sim\cos(\delta\phi/2)$ form, and for the strong barrier limit, for which it takes a $\sim\sin(\delta\phi)$ form characterizing the Josephson regime. In particular, we derived an expression for the Josephson critical current as a function of the link parameters. 

Finally, we found two kinds of multivalued current-phase diagrams which we show, by means of an analytical model, to appear due to the competition between a hydrodynamic and a nonlinear-dispersive part of the flow, which can coexist only for barriers wide enough to accomodate a soliton inside. The first kind, % allowing for phase differences larger than $\pi$
showing two different positive currents at same phase difference, has recently been experimentally demonstrated with $^4${\rm He} and proven to be connected to the appearance of phase slippage in the AC Josephson dynamics. The second one, new to our knowledge, can % involve phase differences always smaller than $\pi$ though being multivalued
instead allow for three different positive currents corresponding to the same phase difference. We believe that this new kind of hysteresis in the current-phase relation can be associated with new features emerging in the Josephson dynamics, which will be studied elsewhere.

\begin{acknowledgments}
%We wish to acknowledge useful conversations with ....
The Los Alamos National Laboratory is operated by Los Alamos National Security, LLC for the National Nuclear Security Administration of the U.S. Department of Energy under Contract No.~DE-AC52-06NA25396. 
\end{acknowledgments}


\vspace{2.0in}
\begin{thebibliography}{99}

%K. K. Likharev, Rev. Mod. Phys. {\bf 51}, 101 (1979).
\bibitem{barone} A. Barone, G. Paterno,  {\emph Physics and Applications of the Josephson Effect} ( John Wiley and Sons, New York, 1982).

\bibitem{likharev} K. K. Likharev,  {\emph Dynamics of Josephson Junctions and Circuits} (Gordon and Breach, New York, 1986).

\bibitem{packard_RMP} R. E. Packard, Rev. Mod. Phys. {\bf 70}, 641 (1998).

\bibitem{varoquaux} O. Avenel and E. Varoquaux, Phys. Rev. Lett. {\bf 55}, 2704 (1985). 

\bibitem{packard} E. Hoskinson, Y. Sato, I. Hahn, and R. E. Packard, Nat. Phys. \textbf{2}, 23 (2006).

\bibitem{baratoff} A. Baratoff, J. A. Blackburn, and B. B. Schwartz,
Phys. Rev. Lett. {\bf 25}, 1096 (1970).

\bibitem{smerzi} A. Smerzi, S. Fantoni, S. Giovanazzi, and S. R. Shenoy, Phys. Rev. Lett. {\bf 79}, 4950 (1997).

\bibitem{milburn} G. J. Milburn, J. Corney, E. M. Wright, and D. F. Walls, Phys. Rev. A {\bf 55}, 4318 (1997).

\bibitem{leggett} I. Zapata, F. Sols, and A. J. Leggett, Phys. Rev. A {\bf 57}, R28 (1998).

\bibitem{bergeman} D. Ananikian and T. Bergeman, Phys. Rev. A {\bf 73}, 013604 (2006).
%\bibitem{giovanazzi} S. Giovanazzi, A. Smerzi, and S. Fantoni, Phys. Rev. Lett. {\bf 84}, 4521 (2000).

\bibitem{chiofalo} M. L. Chiofalo, and M. P. Tosi, Europhys. Lett. {\bf 56}, 326 (2001).

\bibitem{inguscio} F. S. Cataliotti, {\it et al.}, Science {\bf 293}, 843 (2001).

\bibitem{oberthaler} Albiez M., {\it et al.}, Phys. Rev. Lett. {\bf 95}, 010402 (2005).

\bibitem{steinhauer} S. Levy, {\it et al.}, Nature {\bf 449}, 579 (2007).

\bibitem{strinati} A. Spuntarelli, P. Pieri, and G. C. Strinati, Phys. Rev. Lett. {\bf 99}, 040401 (2007).

\bibitem{watanabe} G. Watanabe, {\it et al.}, Phys. Rev. A {\bf 80}, 053602 (2009).

\bibitem{ancilotto} F. Ancilotto, L. Salasnich, and F. Toigo, Phys. Rev. A {\bf 79}, 033627 (2009).

\bibitem{dalfovo} F. Dalfovo, S. Giorgini, L. Pitaevskii, and S. Stringari,
Rev. Mod. Phys. {\bf 71}, 463 (1999).

\bibitem{leboeuf} P. Leboeuf, and N. Pavloff, Phys. Rev. A {\bf 64}, 033602 (2001).

\bibitem{stability} The study of the stability of the two solutions is beyond the scope of the present paper. This issue has been addressed in \cite{hakim}.%  One solution was shown to be unstable while the other was regarded as stable. 

\bibitem{mamaladze} Yu.~G. Mamaladze and O.~D. Che\u{i}shvili,
Zh. Eksp. Teor. Fiz. {\bf 50}, 169 (1966)
[Sov. Phys. JETP {\bf 23}, 112 (1966)]; J. S. Langer, and V. Ambegaokar, Phys. Rev. {\bf 164}, 498 (1967); B. T. Seaman, L. D. Carr, and M. J. Holland, Phys. Rev. A {\bf 71}, 033609 (2005); W. D. Li, Phys. Rev. A {\bf 74}, 063612 (2006).

\bibitem{solutionkind} In general, the two solutions can both correspond to a positive $\Delta$, both to a negative $\Delta$, or one to a positive and the other to a negative $\Delta$.

\bibitem{hakim} V. Hakim, Phys. Rev. E {\bf 55}, 2835 (1997).

\bibitem{lesz} A.~M. Leszczyszyn, G.~A. El, Yu.~G. Gladush, and A.~M. Kamchatnov,
Phys. Rev. A {\bf 79}, 063608 (2009).

% \bibitem{pavloff} N. Pavloff, Phys. Rev. A {\bf 66}, 013610 (2002).

%\bibitem{carr} B. T. Seaman, L. D. Carr, and M. J. Holland  Phys. Rev. A {\bf 71}, 033609 (2005).

\bibitem{deltaphisol} By further approximating $n_2(d)\simeq n_0$ we get $\delta\phi_{\rm sol}=2\arccos(j/\sqrt{gn_0^3/m})$. Since $n_0$ is a monotonically decreasing function of $j$, the argument of $\arccos$ increases for increasing current. Therefore, starting from $\pi$ at $j=0$, $\delta\phi_{\rm sol}$ decreases monotonically with $j$.

\bibitem{closetotip} Close to the critical point $(\delta\phi_c,j_c)$, and for every current-phase relation, the lower branch always has a phase decreasing with increasing current, up to the critical point itself, at which it meets the upper branch. In particular, for reentrant diagrams, this means that the dispersive part of the flow always dominates over the hydrodynamic part sufficiently close to the critical point.

\bibitem{deaver}  B. S. Deaver, and J.M. Pierce, Phys. Lett. A {\bf 38}, 81 (1972). 

\end{thebibliography}
\end{document}